\documentclass[aps,prd,superscriptaddress,showpacs,preprintnumbers,10pt]{revtex4}
\usepackage[dvips]{graphicx,epsfig,color}
\usepackage{wrapfig,rotating}
\usepackage{amssymb,amsmath,array}

\usepackage{color}

\newcommand{\be}{\begin{equation}}
\newcommand{\ee}{\end{equation}}
\newcommand{\bq}{\begin{eqnarray}}
\newcommand{\eq}{\end{eqnarray}}

\newcommand{\D}{\mathrm{d}}
\newcommand{\E}{\mathrm{e}}

\def\Vec#1{\mathpalette{\VVec}{#1}}                  
\def\VVec#1#2{\mbox{\boldmath$#1#2$\unboldmath}}

\def\anti#1{\mathpalette{\@anti}{#1}#1}
\def\@anti#1#2{\sbox0{$#1#2$}
  \makebox[0pt][l]{$#1\kern.30\ht0\overline{\kern-.35\ht0\phantom{#2}}$}}

\begin{document}

\title{A direct extraction of the Sivers distributions \\
from spin asymmetries in pion and kaon leptoproduction}

\author{Anna~Martin}

\affiliation{Dipartimento di Fisica, Universit{\`a}
degli Studi di Trieste; \\
INFN, Sezione di Trieste, 34127 Trieste, Italy}

\author{Franco~Bradamante} 

\affiliation{INFN, Sezione di Trieste, 34127 Trieste, Italy}

\author{Vincenzo~Barone}

\affiliation{Di.S.I.T., Universit{\`a} del Piemonte
Orientale ``A. Avogadro'', 15121 Alessandria, Italy; \\
INFN, Sezione di Torino,  10125 Torino, Italy}

\begin{abstract}
We present a point-by-point determination of the Sivers 
distributions from hadron leptoproduction data. 
The method, which relies on some simple
 assumptions, 
is based on the combined analysis of proton and deuteron observables.  
We make use of the
single-spin asymmetries measured by COMPASS 
in semi-inclusive deep inelastic scattering 
of 160 GeV muons on transversely polarized proton and deuteron targets.

\end{abstract} 


\maketitle

\section{Introduction}

One of the most important achievements of hadronic physics in the past decades 
has been the discovery of significant single-spin asymmetries in leptoproduction 
of hadrons from a transversely polarized target, $\ell N^{\uparrow} \rightarrow 
\ell' h X$ (for reviews, see e.g. \cite{bbm,Aidala:2013,EPJA}).  
One of these asymmetries is associated with a characteristic angular 
modulation of the cross section 
and originates from a correlation between the transverse spin of the 
nucleon and the transverse momentum of quarks, described 
by a leading--twist transverse--momentum dependent distribution (TMD),  
the so-called Sivers function $f_{1T}^{\perp}$ \cite{Sivers:1990,Sivers:1991,Brodsky:2002,Collins:2002}.

The Sivers asymmetry
has been experimentally observed by the HERMES and COMPASS 
collaborations in the case of pion and kaon production
\cite{Airapetian:2005,Alexakhin:2005,Ageev:2006da,%
Airapetian:2009,Alekseev:2009,Alekseev:2010rw,Adolph:2012,Adolph:2015}.
More recently, data on pion production on a transversely polarized $^3He$ target 
have been made available by the Hall A Collaboration at JLab \cite{Qian:2011}. 
Many phenomenological studies of these measurements are available 
in the literature  
\cite{Efremov:2003,Efremov:2005,Collins:2005,Vogelsang:2005,Anselmino:2005a,Anselmino:2005b,%
Anselmino:2006,Anselmino:2009,Anselmino:2012,Sun:2013,Echevarria:2014}. In most analyses 
the Sivers distributions are extracted by fitting the data with a given functional form 
for the dependence of $f_{1T}^{\perp}$ on the Bjorken $x$ variable and 
on the quark intrinsic transverse momentum $k_T^2$. 
Here we adopt a different and simpler approach, similar  
to the one successfully used for the Collins asymmetries in a previous paper of ours 
\cite{Martin:2015}. 
The COMPASS measurements
with proton and deuteron targets in the same kinematics allow 
to perform a point-by-point extraction 
of the Sivers distributions directly from the data, 
by properly combining the various asymmetries.  Although we use 
a Gaussian form for the TMD's in order 
to factorize them from the fragmentation functions, our extraction 
is essentially parameter-free.  
In particular, it does not require any specific assumption about the average values of the 
transverse momenta of quarks. We obtain the Sivers valence distributions 
both in the case of pion production and in the case of kaon production, 
and we show that they are compatible with each other.   

The plan of the paper is the following. In Sec.~II we present the general formalism 
and write the asymmetries for pion and kaon production, showing that 
some combinations of them directly provide the Sivers distributions. In Sec.~III 
we extract the valence and sea Sivers distributions from the asymmetries. Finally, Sec.~IV 
contains some concluding remarks.

\section{Sivers asymmetries}

\subsection{General formulas}

The process we will be considering is semi-inclusive DIS (SIDIS) 
with a transversely polarized target, $\ell N^{\uparrow} \rightarrow 
\ell` h X$.   
We denote by $\Vec P_h$ and $M_h$ the momentum and the mass, respectively, of the produced hadron. 
Conventionally, all azimuthal angles are referred to the lepton scattering plane,  
in a reference system in which the $z$ axis is the virtual photon direction, 
while the $x$ axis is directed along the transverse momentum of the outgoing lepton: 
$\phi_h$ is the azimuthal angle of $\Vec P_h$, $\phi_S$ is 
the azimuthal angle of the nucleon spin vector $\Vec S_{\perp}$.  
The transverse momenta are defined as follows: $\Vec k_T$ is the transverse momentum of the 
quark inside the nucleon,  $\Vec p_T$ is the transverse momentum 
of the hadron with respect to the direction of the fragmenting quark, $\Vec P_{h \perp}$ 
is the measurable transverse momentum of the produced hadron 
with respect to the $z$ axis.

The Sivers term in the cross section, which couples  
 the distribution $f_{1T}^{\perp} (x, k_T^2, Q^2)$ to the 
transverse-momentum dependent unpolarized fragmentation 
function $D_1(z, p_T^2, Q^2)$,  is characterized by 
a $\sin (\phi_h - \phi_S)$ modulation. 
 The corresponding asymmetry is \cite{Mulders:1996,Boer:1998}
\be
A_h(x, z, Q^2) = 
\frac{\sum_a e_a^2 x \int \D^2 \Vec P_{h \perp} \, 
\mathcal{C} \left [ \frac{\Vec P_{h \perp} \cdot \Vec k_T}{M P_{h \perp}} 
\, f_{1T}^{\perp} \, D_1 \right ]}{\sum_a e_a^2 x \int \D^2 \Vec P_{h \perp} \, 
\mathcal{C} \, \left [ f_1 D_1 \right ]},  
\label{sivers}
\ee 
where 
the convolution $\mathcal{C}$ is defined as 
\bq
\mathcal{C} \, [w f D] &=& 
  \int \D^2 \Vec k_T \int \D^2 \Vec p_T
\, \delta^2 ( z \Vec k_T + \Vec p_T - \Vec P_{h \perp}) \nonumber \\
& & \times \,  w(\Vec k_T, \Vec p_T) 
\, f^a (x,  k_T^2, Q^2) D^a (z,  p_T^2, Q^2)\,. 
\label{convol} 
\eq

If we adopt a Gaussian model for 
the transverse-momentum dependent distribution and fragmentation functions:  
\bq
&& f_1(x, k_T^2, Q^2) = f_1(x, Q^2) \, \frac{\E^{- k_T^2/\langle k_T^2 \rangle}}{\pi 
\langle k_T^2 \rangle}\,,
\label{tmd_gauss1} \\
& & f_{1T}^{\perp}(x, k_T^2, Q^2) = f_{1T}^{\perp}(x, Q^2) \, \frac{\E^{- k_T^2/\langle k_T^2 \rangle_S}}{\pi 
\langle k_T^2 \rangle_S}\,,
\label{tmd_gauss2} 
\\
& & 
D_1(z, p_T^2, Q^2)  = D_1(z, Q^2) \, \frac{\E^{- p_T^2/\langle p_T^2 \rangle}}{\pi 
\langle p_T^2 \rangle}\,,  
\label{tmd_gauss3} 
\eq
the Sivers asymmetry (\ref{sivers}) takes the form \cite{Efremov:2003,Efremov:2005,Boer:1998} 
\be
A_h(x,z, Q^2) 
= G \,\,    \frac{\sum_{q,\bar{q}} e_q^2 x f_{1T}^{\perp (1) q}(x, Q^2) z D_{1q}(z, Q^2)}
{\sum_{q,\bar{q}} e_q^2 x f_1^q(x, Q^2)  D_{1q}(z, Q^2)}. 
\label{sivers22}
\ee
Here the first $k_T^2$ moment of the Sivers function is defined as   
\be
f_{1T}^{\perp (1)}(x, Q^2) \equiv \int \D^2  \Vec k_T 
\, \frac{k_T^2}{2 M^2} \,  
\, f_{1T}^{\perp}(x, k_T^2, Q^2) \,,   
\label{k_T_moment}
\ee
and $D_1(z, Q^2)$ is the fragmentation function integrated over the transverse momentum. 
The $G$ factor, resulting from the Gaussian integrations,  is given by \cite{Efremov:2003,Efremov:2005}
\be
G  =  \frac{\sqrt{\pi} M}{\sqrt{\langle p_T^2 \rangle + 
z^2 \langle k_T^2 \rangle_S}},  
\ee
where $\langle k_T^2 \rangle_S$ is the width of the Sivers distribution. 
In the Gaussian model the average transverse momentum of the produced hadrons is 
\be
\langle P_{h \perp} \rangle = \frac{\sqrt{\pi}}{2} 
\sqrt{\langle p_T^2 \rangle + z^2 \langle k_T^2 \rangle}\,, 
\ee
where $\langle k_T^2 \rangle$ is the width of the unpolarized $f_1$ distribution. 
The positivity bound for the Sivers function implies that $\langle k_T^2 \rangle_S$
must be smaller than $\langle k_T^2 \rangle$, but with an error which is well within 
the overall (experimental + model) uncertainties, we can 
identify $G$ with 
\be
G \simeq \frac{\pi M}{2 \langle P_{h \perp} \rangle}. 
\label{gfactor}
\ee
$\langle P_{h \perp} \rangle$ is experimentally 
found to have a very mild dependence on $x$ and $z$. For simplicity 
we take its value averaged over $z$, 
so that the $G$ factor used in our calculations will slightly depend on $x$ only.

Since our aim is to extract the $k_T^2$ moment of the Sivers distribution, 
we integrate over $z$ 
\be
\widetilde{D}_{1}(Q^2) = 
\int \D z \, D_{1} (z, Q^2)\,, 
\;\;\;\;
 \widetilde{D}_{1}^{(1)}(Q^2) = 
\int \D z \, z D_{1}(z, Q^2)\,, 
\ee
and consider the integrated asymmetry:   
\be
A_{h}(x, Q^2) = G \,  
\frac{\sum_{q,\bar{q}} e_q^2 x f_{1T}^{\perp (1) q}(x, Q^2)  \widetilde{D}_{1q}^{(1)}(Q^2)}
{\sum_{q,\bar{q}} e_q^2 x f_1^q(x, Q^2)  \widetilde{D}_{1q} (Q^2)}. 
\label{eq:sivers}
\ee

\subsection{Pion production}

It is convenient to 
distinguish favored and unfavored fragmentation functions. For pions they 
are defined as 
\begin{eqnarray}
& & D_{1, {\rm fav}}^{\pi} \equiv D_{1u}^{\pi^+} = D_{1d}^{\pi^-} = D_{1\bar{u}}^{\pi^-} 
= D_{1\bar{d}}^{\pi^+} \label{eq:favdis1} \\
& & D_{1, {\rm unf}}^{\pi} \equiv D_{1u}^{\pi^-} = D_{1d}^{\pi^+} = D_{1\bar{u}}^{\pi^+} 
= D_{1\bar{d}}^{\pi^-}\,. 
\label{eq:favdis2}
\end{eqnarray}
As for the strange sector, we take 
\be
D_{1s}^{\pi^\pm} = D_{1 \bar s}^{\pi^\pm} = N \, D_{1, {\rm unf}}^{\pi}\,,  
\ee
where $N$ is a constant coefficient. In the fragmentation function parametrization 
of Ref.~\cite{deflorian}, $N$ is found to be 0.83.

The denominators of the 
asymmetries $\sum_{q, \bar q} e_q^2 x f_1^q \widetilde{D}_{1q}$,
for a proton and a deuteron target ($p, d$) and for charged pions, 
multiplied by 9, are given by (we ignore the charm components of the 
distribution functions, which are negligible in the kinematic region 
we are interested in)
\begin{eqnarray}
\hspace{-0.5cm} & & p, \pi^+: 
\;\;\;\;
 x \,  [ 4 (f^{u}_1 + \beta_{\pi} f^{\bar{u}}_1) +
( \beta_{\pi} f^{d}_1 + f^{\bar{d}}_1 )  + N \beta_{\pi} (f_1^s + f_1^{\bar s})] \,
 \widetilde{D}_{1, {\rm fav}}^{\pi} 
\equiv x f_p^{\pi^+} \, \widetilde{D}_{1, {\rm fav}}^{\pi}, 
\label{p+} \\
\hspace{-0.5cm} & & d, \pi^+: 
\;\;\;\; 
  x \,  [ (4 + \beta_{\pi}) (f^{u}_1 + f^{d}_1 ) 
+ (1 + 4 \beta_{\pi}) (f^{\bar{u}}_1 + f^{\bar{d}}_1) + 2 N \beta_{\pi} (f_1^s + f_1^{\bar s})]\,
 \widetilde{D}_{1, {\rm fav}}^{\pi}
\equiv x f_d^{\pi^+} \, \widetilde{D}_{1, {\rm fav}}^{\pi},  
\label{d+} \\
\hspace{-0.5 cm}
& & p, \pi^-: \;\;\;\;
 x \, [4 (\beta_{\pi} f^{u}_1 + f^{\bar{u}}_1) +
( f^{d}_1 + \beta_{\pi} f^{\bar{d}}_1 ) + 
N \beta_{\pi} (f_1^s + f_1^{\bar s})] \, \widetilde{D}_{1, {\rm fav}}^{\pi}
\equiv x f_p^{\pi^-} \, \widetilde{D}_{1, {\rm fav}}^{\pi} ,  
\label{p-} \\
\hspace{-0.5cm} 
& & d, \pi^-: \;\;\;\;
x \, [( 1 + 4 \beta_{\pi}) (f^{u}_1 + f^{d}_1 ) +
( 4 + \beta_{\pi}) (f^{\bar{u}}_1 + f^{\bar{d}}_1) + 2 
N \beta_{\pi} (f_1^s + f_1^{\bar s}) ]
\, \widetilde{D}_{1, {\rm fav}}^{\pi} 
\equiv x f_d^{\pi^-} \, \widetilde{D}_{1,  {\rm fav}}^{\pi}, 
\label{d-}
\end{eqnarray}
with
\be
\beta_{\pi} (Q^2) = \frac{\widetilde{D}_{1, {\rm unf}}^{\pi}(Q^2)}{\widetilde{D}_{1, {\rm fav}}^{\pi}
 (Q^2)}. 
\label{betaq}
\ee
Similar expressions can be written for the numerator of Eq.~(\ref{eq:sivers}), 
$\sum_{q, \bar q} e_q^2 x f_{1T}^{\perp (1) q} \widetilde{D}_{1q}^{(1)}$, 
with the replacements
$\widetilde{D}_1 \to \widetilde{D}_1^{(1)}$, $f_1 \to f_{1T}^{\perp (1)}$, and $\beta_{\pi} \to 
\beta_{\pi}^{(1)}$,  where
\be
\beta_{\pi}^{(1)}(Q^2) = \frac{\widetilde{D}_{1, {\rm unf}}^{\pi (1)}(Q^2)}
{\widetilde{D}_{1, {\rm fav}}^{\pi (1)}(Q^2)}. 
\label{alphaq}
\ee

Introducing the ratio of the first to the zeroth moment of the fragmentation functions, 
\be
\rho_{\pi} (Q^2) = \frac{\widetilde{D}_{1, {\rm fav}}^{\pi (1)}(Q^2)}
{\widetilde{D}_{1, {\rm fav}}^{\pi} (Q^2)}\,,
\label{mom_ratio}
\ee 
we find for the pion asymmetries with a proton target
\begin{eqnarray}
& & A^{\pi^+}_{p}= G \, 
\rho_{\pi} \, 
\frac{4 (f_{1T}^{\perp (1) u} + \beta_{\pi}^{(1)} f_{1T}^{\perp (1) \bar{u}}) +
 ( \beta_{\pi}^{(1)} f_{1T}^{\perp (1) d} + f_{1T}^{\perp (1) \bar{d}} ) 
+ N \beta_{\pi}^{(1)} (f_{1T}^{\perp (1) s} + f_{1T}^{\perp (1) \bar{s}} ) }{ f_p^{\pi^+}}, 
\label{asym_p+} \\
& & A^{\pi^-}_{p}= G \, 
\rho_{\pi} \, 
 \frac{4  (\beta_{\pi}^{(1)} f_{1T}^{\perp (1) u} + f_{1T}^{\perp (1) \bar{u}}) +
 (f_{1T}^{\perp (1) d} + \beta_{\pi}^{(1)}  f_{1T}^{\perp (1) \bar{d}} ) 
+ N \beta_{\pi}^{(1)} (f_{1T}^{\perp (1) s} + f_{1T}^{\perp (1) \bar{s}} )}{ f_p^{\pi^-}}, 
\label{asym_p-}
\end{eqnarray} 
and for the deuteron target 
\begin{eqnarray}
& & A^{\pi^+}_{d} = G \, 
\rho_{\pi} \, 
 \frac{(4 + \beta_{\pi}^{(1)})  (f_{1T}^{\perp (1) u} + f_{1T}^{\perp (1) d} )  +
(1 + 4 \beta_{\pi}^{(1)})  (f_{1T}^{\perp (1) \bar{u}} + f_{1T}^{\perp (1) \bar{d}})
+ 2 N \beta_{\pi}^{(1)} (f_{1T}^{\perp (1) s} + f_{1T}^{\perp (1) \bar{s}} ) }{ f_d^{\pi^+}}, 
\label{asym_d+} \\
& & A^{\pi^-}_{d} = G\, 
\rho_{\pi} \, 
\frac{(1 + 4 \beta_{\pi}^{(1)})  (f_{1T}^{\perp (1) u} + f_{1T}^{\perp (1) d} )  +
(4 + \beta_{\pi}^{(1)})  (f_{1T}^{\perp (1) \bar{u}} + f_{1T}^{\perp (1) \bar{d}}) 
+ 2 N \beta_{\pi}^{(1)} (f_{1T}^{\perp (1) s} + f_{1T}^{\perp (1) \bar{s}} ) }{ f_d^{\pi^-}} .
\label{asym_d-}
\end{eqnarray}
The combinations 
\begin{eqnarray}
& &  f_p^{\pi^+} A^{\pi^+}_{p} -  f_p^{\pi^-} A^{\pi^-}_{p} =
G \, \rho_{\pi}
(1 - \beta_{\pi}^{(1)}) (4 f_{1T}^{\perp (1) u_v} - f_{1T}^{\perp (1) d_v}  )
\label{p_difference} \\
& &  f_d^{\pi^+} A^{\pi^+}_{d} -  f_d^{\pi^-} A^{\pi^-}_{d} =
3 G \, \rho_{\pi}  
 (1 - \beta_{\pi}^{(1)}) (f_{1T}^{\perp (1) u_v} + f_{1T}^{\perp (1) d_v}  )
\label{d_difference}
\end{eqnarray}
select the valence Sivers distributions.
From eqs.~(\ref{p_difference}, \ref{d_difference}), we get 
the valence distributions for $u$ and $d$ quarks separately: 
\bq
& & x f_{1T}^{\perp (1) u_v} = \frac{1}{5 G \rho_{\pi} (1 - \beta_{\pi}^{(1)})} 
\left [ ( x f_p^{\pi^+} A_p^{\pi^+} - x f_p^{\pi^-} A_p^{\pi^-}) + 
\frac{1}{3}  ( x f_d^{\pi^+} A_d^{\pi^+} - x f_d^{\pi^-} A_d^{\pi^-})  \right ] \,, 
\label{uval_siv}
\\
& & x f_{1T}^{\perp (1) d_v} = \frac{1}{5 G \rho_{\pi} (1 - \beta_{\pi}^{(1)})} 
\left [ \frac{4}{3}  ( x f_d^{\pi^+} A_d^{\pi^+} - x f_d^{\pi^-} A_d^{\pi^-}) -  
(x f_p^{\pi^+} A_p^{\pi^+} - x f_p^{\pi^-} A_p^{\pi^-}) \right ]\,.
\label{dval_siv}
\eq

A particular combination of proton and deuteron asymmetries selects the sea component 
of the Sivers function, namely
\bq
 x f_{1T}^{\perp (1) \bar u} - x f_{1T}^{\perp (1) \bar d} &=& \frac{1}{15 G \rho_{\pi} 
\left (1 - \beta_{\pi}^{(1) 2} \right )} 
\left [ 2 (1 - 4 \beta_{\pi}^{(1)}) x f_p^{\pi^+} A_p^{\pi^+} 
+ 2 (4 - \beta_{\pi}^{(1)}) x f_p^{\pi^-} A_p^{\pi^-} 
\right. \nonumber \\
& & - \left. (1 - 4 \beta_{\pi}^{(1)}) x f_d^{\pi^+} A_d^{\pi^+} 
- (4 - \beta_{\pi}^{(1)}) x f_d^{\pi^-} A_d^{\pi^-})  \right ] \,.  
\label{ubar-dbar}
\eq

In Sec.~\ref{extraction} we will apply Eqs.~(\ref{uval_siv}, \ref{dval_siv}, \ref{ubar-dbar}) 
to extract the Sivers valence and sea distributions.

\subsection{Kaon production}

In the case of kaons we have two favored fragmentation 
functions, which largely differ from each other:  
\bq
& &  D_{1, {\rm fav}}^K \equiv D_{1u}^{K^+}  = D_{1\bar{u}}^{K^-} \\
\label{eq:fav1_kaons} 
& & D_{1, {\rm fav}}'^K \equiv D_{1\bar{s}}^{K^+} = D_{1 s}^{K^-} 
\label{eq:fav2_kaons} 
\eq
Since it is more difficult to excite from the vacuum a heavy $s \bar s $ pair 
than a light $u \bar u$ pair, $D_{1, {\rm fav}}'^K$ is expected 
to be (and in fact is) much larger than $D_{1, {\rm fav}}^K$. The unfavored 
fragmentation functions are defined as usual: 
\be
 D_{1, {\rm unf}}^K \equiv D_{1d}^{K^\pm} = D_{1\bar{d}}^{K^\pm} = D_{1\bar{u}}^{K^+} 
= D_{1 u}^{K^-}= D_{1s}^{K^+} = D_{1 \bar s}^{K^-}\,.     
\label{eq:unfav_kaons}
\ee

Proceeding as before, the denominators of the 
asymmetries $\sum_{q, \bar q} e_q^2 x f_1^q \widetilde{D}_{1q}$,
for kaon production from a proton and a deuteron target ($p, d$), 
multiplied by 9, are given by
\begin{eqnarray}
\hspace{-0.5cm} & & p, K^+: 
\;\;\;\;
 x \,  [ 4 (f^{u}_1 + \beta_K f^{\bar{u}}_1) +
\beta_K (f^{d}_1 + f^{\bar{d}}_1 )  + (\beta_K f_1^s + \gamma_K f_1^{\bar s})] \,
 \widetilde{D}_{1, {\rm fav}}^K 
\equiv x f_p^{K^+} \, \widetilde{D}_{1, {\rm fav}}^K, 
\label{pK+} \\
\hspace{-0.5cm} & & d, K^+: 
\;\;\;\; 
  x \,  [ (4 + \beta_K) (f^{u}_1 + f^{d}_1 ) 
+  5 \beta_K (f^{\bar{u}}_1 + f^{\bar{d}}_1) + 2 (\beta_K f_1^s + \gamma_K f_1^{\bar s})]\,
 \widetilde{D}_{1, {\rm fav}}^K
\equiv x f_d^{K^+} \, \widetilde{D}_{1, {\rm fav}}^K,  
\label{dK+} \\
\hspace{-0.5 cm}
& & p, K^-: \;\;\;\;
 x \, [4 (\beta_K f^{u}_1 + f^{\bar{u}}_1) +
\beta_K ( f^{d}_1 +  f^{\bar{d}}_1 ) + 
(\gamma_K f_1^s + \beta_K f_1^{\bar s})] \, \widetilde{D}_{1, {\rm fav}}^K
\equiv x f_p^{K^-} \, \widetilde{D}_{1, {\rm fav}}^K ,  
\label{pK-} \\
\hspace{-0.5cm} 
& & d, K^-: \;\;\;\;
x \, [5 \beta_K (f^{u}_1 + f^{d}_1 ) +
( 4 + \beta_K) (f^{\bar{u}}_1 + f^{\bar{d}}_1) + 2 
 (\gamma_K f_1^s + \beta_K f_1^{\bar s}) ]
\, \widetilde{D}_{1, {\rm fav}}^K 
\equiv x f_d^{K^-} \, \widetilde{D}_{1,  {\rm fav}}^K, 
\label{dK-}
\end{eqnarray}
where $\beta_K$ and $\gamma_K$ are defined as 
\be
\beta_{K} (Q^2) = \frac{\widetilde{D}_{1, {\rm unf}}^{K}(Q^2)}{\widetilde{D}_{1, {\rm fav}}^{K}
 (Q^2)}, 
\;\;\;\;\;
\gamma_{K} (Q^2) = \frac{\widetilde{D}_{1, {\rm fav}}'^{K}(Q^2)}{\widetilde{D}_{1, {\rm fav}}^{K}
 (Q^2)}, 
\label{beta_gamma_K}
\ee

For the numerator of Eq.~(\ref{eq:sivers}), 
$\sum_{q, \bar q} e_q^2 x f_{1T}^{\perp (1) q} \widetilde{D}_{1q}^{(1)}$, 
one can write similar expressions with the replacements
$\widetilde{D}_1 \to \widetilde{D}_1^{(1)}$, $f_1 \to f_{1T}^{\perp (1)}$, $\beta_K \to 
\beta_K^{(1)}$ and $\gamma_K \to \gamma_K^{(1)}$,  where $\beta_K^{(1)}$ and 
$\gamma_K^{(1)}$ are given by
\be
\beta_{K}^{(1)} (Q^2) = \frac{\widetilde{D}_{1, {\rm unf}}^{K (1)}(Q^2)}{\widetilde{D}_{1, 
{\rm fav}}^{K (1)}
 (Q^2)}, 
\;\;\;\;\;
\gamma_{K}^{(1)} (Q^2) = \frac{\widetilde{D}_{1, {\rm fav}}'^{K (1)}(Q^2)}{\widetilde{D}_{1, 
{\rm fav}}^{K (1)}
 (Q^2)}, 
\label{beta_gamma_K(1)}
\ee

The resulting $K^{\pm}$ Sivers asymmetries are: 
\begin{eqnarray}
& & A^{K^+}_{p}=
G \rho_K
\frac{4 (f_{1T}^{\perp (1) u} + \beta_K^{(1)} f_{1T}^{\perp (1) \bar{u}}) +
 \beta_K^{(1)} (f_{1T}^{\perp (1) d} + f_{1T}^{\perp (1) \bar{d}} ) 
+ (\beta_K^{(1)} f_{1T}^{\perp (1) s} + \gamma_K^{(1)} f_{1T}^{\perp (1) \bar{s}} ) }{ f_p^{K^+}}, 
\label{asym_pK+} \\
& & A^{K^-}_{p}=
G \rho_K
 \frac{4  (\beta_K^{(1)} f_{1T}^{\perp (1) u} + f_{1T}^{\perp (1) \bar{u}}) +
 \beta_K^{(1)} (f_{1T}^{\perp (1) d} +   f_{1T}^{\perp (1) \bar{d}} ) 
+ (\gamma_K^{(1)} f_{1T}^{\perp (1) s} + \beta_K^{(1)} f_{1T}^{\perp (1) \bar{s}} )}{ f_p^{K^-}}, 
\label{asym_pK-} \\
& & A^{K^+}_{d} =
G \rho_K
 \frac{(4 + \beta_K^{(1)})  (f_{1T}^{\perp (1) u} + f_{1T}^{\perp (1) d} )  +
5 \beta_K^{(1)}  (f_{1T}^{\perp (1) \bar{u}} + f_{1T}^{\perp (1) \bar{d}})
+ 2  (\beta_K^{(1)} f_{1T}^{\perp (1) s} + \gamma_K^{(1)} f_{1T}^{\perp (1) \bar{s}} ) }{ f_d^{K^+}}, 
\label{asym_dK+} \\
& & A^{K^-}_{d} =
G \rho_K
\frac{5 \beta_K^{(1)}  (f_{1T}^{\perp (1) u} + f_{1T}^{\perp (1) d} )  +
(4 + \beta_K^{(1)})  (f_{1T}^{\perp (1) \bar{u}} + f_{1T}^{\perp (1) \bar{d}}) 
+ 2 (\gamma_K^{(1)} f_{1T}^{\perp (1) s} + \beta_K^{(1)} f_{1T}^{\perp (1) \bar{s}} ) }{ f_d^{K^-}}, 
\label{asym_dK-}
\end{eqnarray}
with the moment ratio $\rho_K$ defined as  
\be
\rho_{K} (Q^2) = \frac{\widetilde{D}_{1, {\rm fav}}^{K (1)}(Q^2)}
{\widetilde{D}_{1, {\rm fav}}^{K} (Q^2)}\,. 
\label{mom_ratio_K}
\ee 

By combining the asymmetries, we get
\bq
& & 
f_p^{K^+} A_p^{K^+} -  f_p^{K^-} A_p^{K^-} = 
G \rho_K \, \left [ 
4 (1 - \beta_K^{(1)}) 
 f_{1T}^{\perp (1) u_v}
+ (\beta_K^{(1)} - \gamma_K^{(1)}) 
(f_{1T}^{\perp (1) s} - f_{1T}^{\perp (1) \bar s})
\right ] 
\label{uval_siv_kaon}
\\
& & 
f_d^{K^+} A_d^{K^+} -  f_d^{K^-} A_d^{K^-} = 
G \rho_K \, \left [ 
4 (1 - \beta_K^{(1)}) 
( f_{1T}^{\perp (1) u_v} + f_{1T}^{\perp (1) d_v})
\right. 
\nonumber \\
& & \hspace{4.5cm} + \left. 2 (\beta_K^{(1)} - \gamma_K^{(1)}) 
(f_{1T}^{\perp (1) s} - f_{1T}^{\perp (1) \bar s})
\right ] . 
\label{udval_siv_kaon}
\eq

In order to extract $u_V$ and $d_V$ separately, we 
assume that the difference of strange sea distributions, $s - \bar s$, 
is negligible. Thus we obtain
\bq
& & x f_{1T}^{\perp (1) u_v} = 
\frac{1}{4 G \rho_{K} (1 - \beta_{K}^{(1)})} 
\left [ ( x f_p^{K^+} A_p^{K^+} - x f_p^{K^-} A_p^{K^-}) \right ]\,, 
\label{uval_siv_K}
\\
& & x f_{1T}^{\perp (1) d_v} = \frac{1}{4 G \rho_{K} (1 - \beta_{K}^{(1)})} 
\left [  ( x f_d^{K^+} A_d^{K^+} - x f_d^{K^-} A_d^{K^-}) -  
(x f_p^{K^+} A_p^{K^+} - x f_p^{K^-} A_p^{K^-}) \right ]\,.
\label{dval_siv_K}
\eq
These two relations will be used to 
extract the valence Sivers distributions from kaon data. 

We note that by using the full set of eight asymmetries 
experimentally measured (proton and deuteron targets, $\pi^{\pm}$ and $K^{\pm}$ productions) 
one could in principle obtain more information on the distributions, 
in particular on the strange and non-strange sea, 
but we prefer to consider only the functions that can be directly determined 
by linear combinations of the asymmetries.

\section{Extraction of the Sivers distributions} 
\label{extraction}

As we have seen,  it is possible 
to obtain directly the valence and the sea components 
of the Sivers function by combining different asymmetries. 
The data we use are from COMPASS measurements of semi-inclusive deep inelastic scattering 
on proton \cite{Adolph:2015} and deuteron targets \cite{Alekseev:2009}. 

In order to extract $f_{1T}^{\perp (1)}$ we need two extra ingredients: 
the unpolarized distribution functions $f_1^q$, which are taken 
from the CTEQ5D global fit \cite{cteq}, and the unpolarized fragmentation 
functions, which are taken from 
the DSS parametrization \cite{deflorian}. 
Notice that in the DSS 
fit of pion fragmentation functions $D_{1u}^{\pi^+}$ is not assumed to be equal to $D_{1 \bar d}^{\pi^+}$, 
but their difference is rather small. Thus, we identify $D_{1, {\rm fav}}$ 
with $(D_{1 u}^{\pi^+} + D_{1 \bar d}^{\pi^+})/2$ as given by DSS.

The normalization of the Sivers distributions is determined 
by the quantity $G=\pi M / 2 \langle P_{h \perp} \rangle$. The  
values of $\langle P_{h \perp} \rangle$, measured by COMPASS, slightly depend on $x$,  
so that $G$ ranges from 2.8 to 3.1 for pions, and from 2.4 to 2.6 for 
kaons.

\begin{figure}[tb]
\centering
\includegraphics[width=0.45\textwidth]{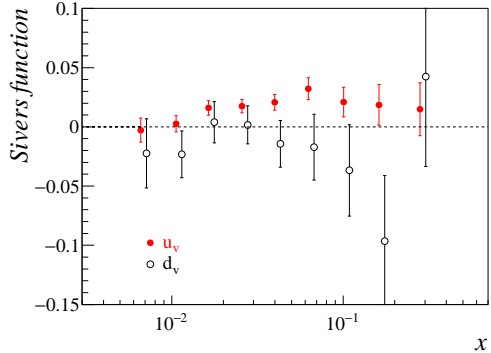}
\caption{The first $k_T^2$ moments of the Sivers valence distributions, $x f_{1T}^{\perp (1) u_v}$ 
(red solid circles)
and $x f_{1T}^{\perp (1) d_v}$ (black open circles),  extracted from 
pion asymmetries.}
\label{fig:uv_dv_pions}
\end{figure}

We can now use Eqs.~(\ref{uval_siv}, \ref{dval_siv}) to extract point-by-point the 
valence Sivers distributions from pion data. The results are tabulated in Table~\ref{tab:valence_pions} 
and displayed in Fig.~\ref{fig:uv_dv_pions}. 
The error bars are computed from the statistical errors of the measured 
asymmetries, and no attempt has 
been made to try to assign a systematic error to the results. 
Notice also that the $x$ points correspond to different $Q^2$ values, ranging from 1.2 GeV$^2$ to
20 GeV$^2$, with an average value $\langle Q^2 \rangle \approx 4$ GeV$^2$.

\begin{table}[tbh]
\begin{center}
\begin{tabular}{|c|c|c|c|}
\hline
$x$ & $Q^2$ (GeV$^2$) & $ x f_{1T}^{\perp (1) u_v}$  & $x f_{1T}^{\perp (1) d_v}$   \\
\hline
 0.007 & 1.22 & -0.003 $\pm$  0.010 &   -0.022 $\pm$  0.029  \\ 
 0.011 & 1.43 &  0.003 $\pm$  0.007 &   -0.023 $\pm$  0.020  \\ 
 0.016 & 1.66 &  0.016 $\pm$  0.006 &    0.004 $\pm$  0.017  \\ 
 0.026 & 1.96 &  0.017 $\pm$  0.006 &    0.002 $\pm$  0.016  \\ 
 0.040 & 2.57 &  0.021 $\pm$  0.007 &   -0.014 $\pm$  0.020  \\ 
 0.063 & 4.01 &  0.032 $\pm$  0.009 &   -0.017 $\pm$  0.028  \\ 
 0.101 & 6.38 &  0.021 $\pm$  0.012 &   -0.037 $\pm$  0.039  \\ 
 0.162 & 9.91 &  0.019 $\pm$  0.017 &   -0.096 $\pm$  0.055  \\ 
 0.281 & 20.23 &  0.015 $\pm$  0.022 &    0.042 $\pm$  0.076  \\ 
\hline
\end{tabular}
\caption{Sivers valence distributions extracted from 
pion asymmetries.}
\label{tab:valence_pions}
\end{center}
\end{table}

The $u_v$ distribution is determined much more precisely   
 than the $d_v$ distribution, due to the fact that
the asymmetry measurements on the proton are considerably more accurate than 
the corresponding ones on the deuteron, in particular in the valence region 
(the COMPASS Collaboration has taken much less data on deuterons 
than on protons). 
Still, 
the $d_v$ Sivers function is reasonably well determined and turns out
to be negative and approximately specular to the 
$u_v$ function. 

\begin{figure}[!ht]
\centering
\includegraphics[width=0.45\textwidth]{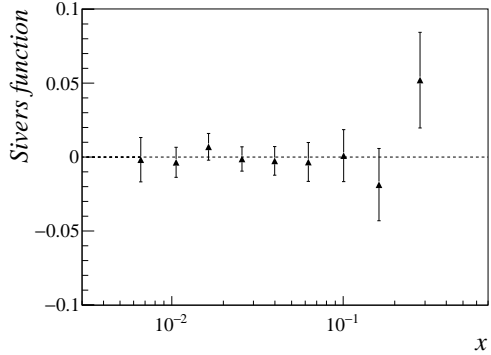}
\caption{The isotriplet Sivers sea $x f_{1T}^{\perp (1) \bar u} - x f_{1T}^{\perp (1) \bar d}$ 
extracted from pion asymmetry data.}
\label{fig:ubmdb}
\end{figure}

Equation~(\ref{ubar-dbar}) allows determining directly the isotriplet $\bar u - \bar d$ component 
of the Sivers sea. The results are shown in Fig.~\ref{fig:ubmdb} and have errors comparable to those 
of the $u_v$ function.
In the large $N_c$ limit, the isotriplet ($\bar u - \bar d$) Sivers combination 
is expected to dominate
over the isosinglet one ($\bar u + \bar d$) \cite{Pobylitsa:2003}, thus the vanishing of 
 $x f_{1T}^{\perp (1) \bar u} - x f_{1T}^{\perp (1) \bar d}$ 
is not due to a cancellation of the two terms, but rather signals 
the smallness of the $\bar u$ and $\bar d$ separately.

\begin{figure}[!ht]
\centering
\includegraphics[width=0.45\textwidth]{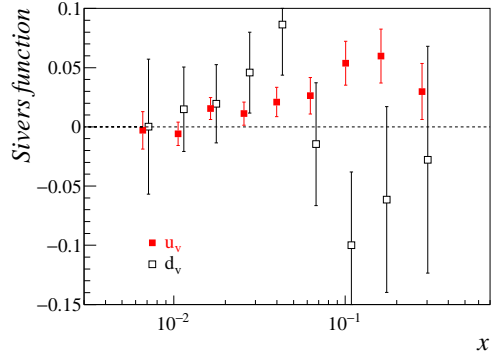}
\caption{The first $k_T^2$ moments of the Sivers valence distributions, $x f_{1T}^{\perp (1) u_v}$ 
(red solid squares)
and $x f_{1T}^{\perp (1) d_v}$ (black open squares),  extracted from 
kaon asymmetries.}
\label{fig:uv_dv_kaons}
\end{figure}

A similar procedure has been applied to extract 
the Sivers functions from the measured kaon asymmetries. The Sivers valence distributions, 
extracted using Eqs.~(\ref{uval_siv_K}, \ref{dval_siv_K}),  
are shown in Table~\ref{tab:valence_kaons} and Fig.~\ref{fig:uv_dv_kaons}.  
Notice that the $Q^2$ values in each $x$ bin are slightly larger 
than for pions. 
Again, the $u_v$ distribution is well determined, whereas in this case 
the $d_v$  distribution is affected by large uncertainties 
and does not exhibit a clear behavior. 
 
\begin{table}[tbh]
\begin{center}
\begin{tabular}{|c|c|c|c|}
\hline
$x$ & $Q^2$ (GeV$^2$) &   $x f_{1T}^{\perp (1) u_v}$  &  $x f_{1T}^{\perp (1) d_v}$   \\
\hline
 0.007 & 1.21 & -0.003 $\pm$  0.016 &   0.000 $\pm$  0.057 \\ 
 0.011 & 1.43 & -0.006 $\pm$  0.010 &   0.015 $\pm$  0.036 \\ 
 0.016 & 1.75 &  0.015 $\pm$  0.009 &   0.020 $\pm$  0.033 \\ 
 0.026 & 2.31 &  0.011 $\pm$  0.010 &   0.046 $\pm$  0.034 \\ 
 0.040 & 3.34 &  0.021 $\pm$  0.012 &   0.086 $\pm$  0.043 \\ 
 0.063 & 5.16 &  0.026 $\pm$  0.015 &  -0.015 $\pm$  0.052 \\ 
 0.101 & 8.01 &  0.054 $\pm$  0.018 &  -0.100 $\pm$  0.062 \\ 
 0.162 & 12.78 &  0.060 $\pm$  0.023 &  -0.061 $\pm$  0.078 \\ 
 0.281 & 26.47 &  0.030 $\pm$  0.024 &  -0.028 $\pm$  0.096 \\ 
\hline
\end{tabular}
\caption{Sivers valence distributions extracted from kaon production.} 
\label{tab:valence_kaons}
\end{center}
\end{table}

\begin{figure}[!ht]
\centering
\includegraphics[width=0.45\textwidth]{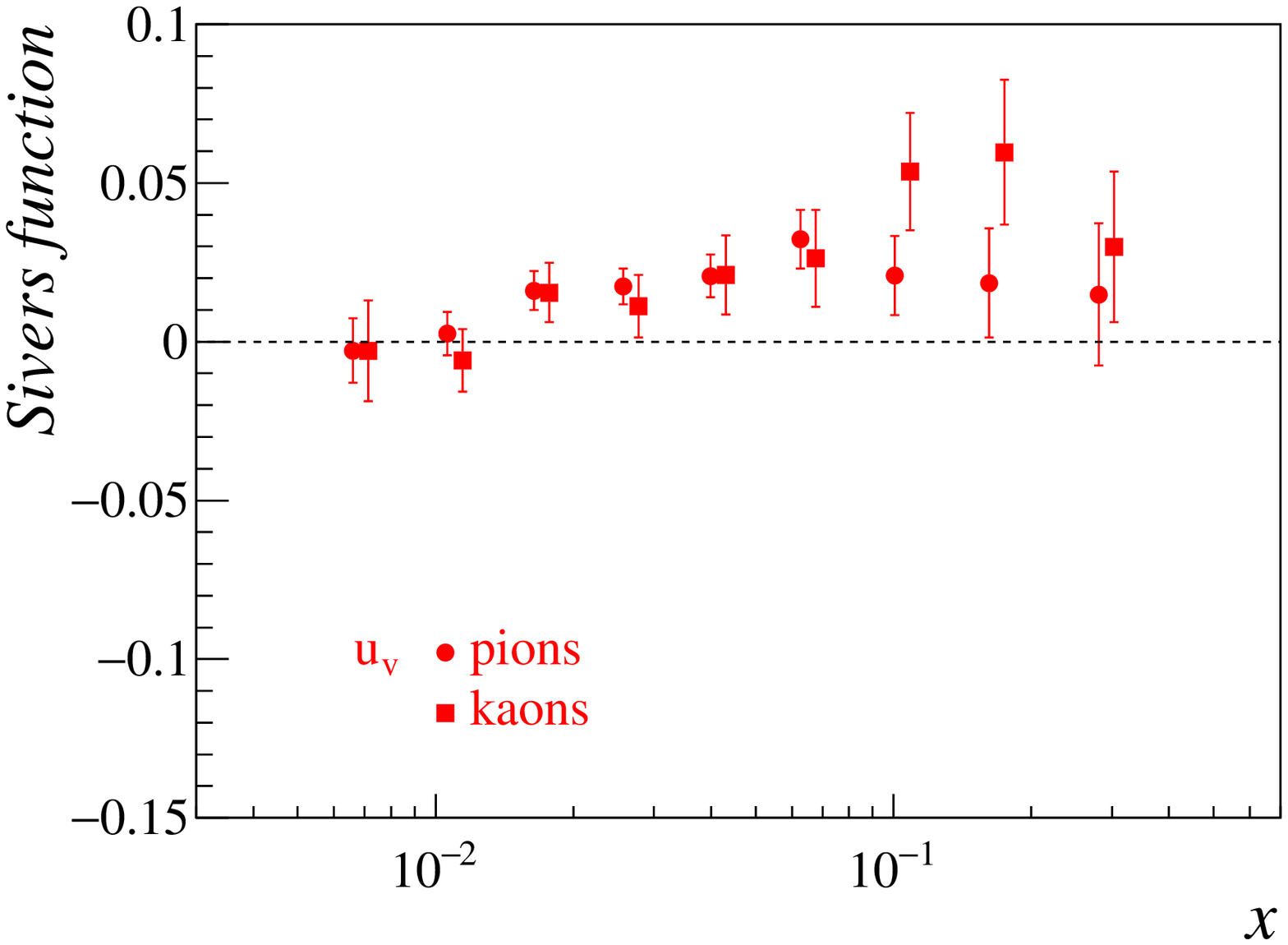}
\includegraphics[width=0.45\textwidth]{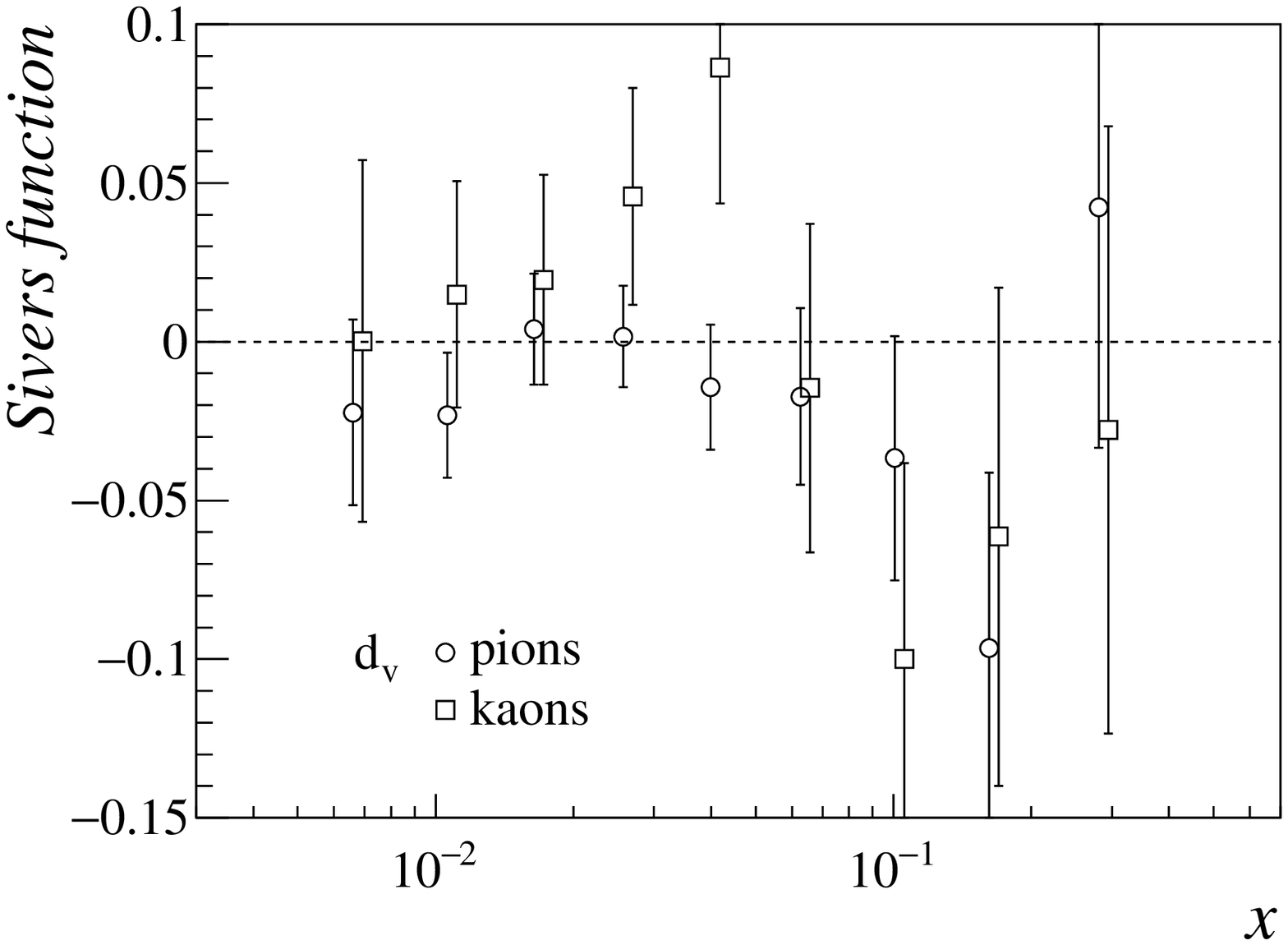}
\caption{Comparison of the first $k_T^2$ moments of the Sivers valence distributions, 
$x f_{1T}^{\perp (1) u_v}$ (left panel) and $x f_{1T}^{\perp (1) d_v}$ (right panel),  
obtained from pion and kaon data.}
\label{fig:val_ud}
\end{figure}

The valence Sivers distributions extracted from 
pion and kaon leptoproduction data are compared in Fig.~\ref{fig:val_ud}.  
In the case of $u_v$ the two sets of points are compatible with 
each other, as they should be, representing the same universal 
property of the target (minor differences in the $Q^2$ values 
of pion and kaon points can be practically ignored). The $d_v$ functions 
are also similar to each other, although affected by much larger errors.

\begin{figure}[!ht]
\centering
\includegraphics[width=0.45\textwidth]{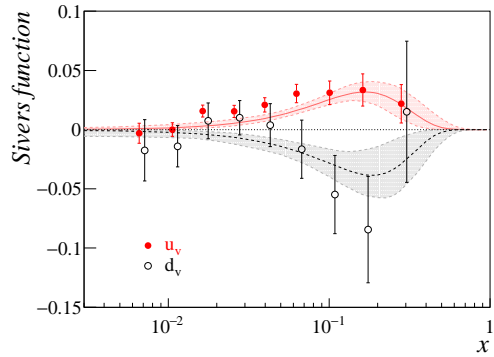}
\caption{The average Sivers distributions  
$x f_{1T}^{\perp (1) u_v}$ (solid points) and $x f_{1T}^{\perp (1) d_v}$ (open points) obtained 
from pion and kaon data.  Our points are compared to the results 
of the fit of Ref.~\cite{Anselmino:2012} for $Q^2 = 4$ GeV$^2$ (central values and uncertainty bands).}
\label{fig:val_average}
\end{figure}

In Fig.~\ref{fig:val_average} we show the weighted averages of the 
Sivers distributions obtained from pion and kaon data. 
For comparison we plot the results (central values 
and uncertainty bands) of the 
fit of Ref.~\cite{Anselmino:2012} based on DGLAP evolution. 
Note that these results 
refer to $Q^2 = 4$ GeV$^2$, the average momentum transfer of 
COMPASS measurements, whereas our points correspond to different 
$Q^2$ values, as explained above. However, except for the first few points at low $x$  
and small momentum transfer, the $Q^2$ evolution 
is not expected to affect the results significantly.  
The TMD evolution  \cite{Aybat:2011,Aybat:2012} has also been applied to the analysis
of the Sivers data, but    
 in this scheme 
the perturbative evolution is driven by a factor which 
cancels out in the asymmetry ratio. Thus, a fit to the asymmetry data 
based on the TMD evolution is not able to constrain the absolute normalization 
of the Sivers distributions (see the discussion in Ref.~\cite{Anselmino:2012}).

\section{Concluding remarks}

In summary, we extracted in a simple and direct way the 
Sivers distributions from the COMPASS measurements of 
pion and kaon leptoproduction on proton and deuteron targets.  
The main assumption we made in order to factorize the Sivers functions from 
the fragmentation functions was the Gaussian behavior in the transverse momenta. 
As expected, the $u_v$ and $d_v$ distributions extracted from the pion data 
are well compatible with the corresponding ones extracted from 
the kaon data, and the final results have been obtained by averaging 
the two partial results. The distributions are 
roughly mirror-symmetric and of similar magnitude, the $u_v$ being positive 
and the $d_v$ negative. They are in good agreement with the 
results of previous fits which assumed a functional form 
for the distributions from the very beginning.  
 
While the $u_v$ distribution is determined with a satisfactory  
accuracy, the $d_v$ distribution is more uncertain. 
To improve its knowledge more data are needed, 
in particular on the deuteron. 
The long-term solution would be the planned Electron Ion 
Collider, but in the near future the 
proposed new COMPASS run on a deuteron target \cite{EPSG} would certainly 
provide new precious information.

Another interesting result from our work is the extraction of the 
Sivers sea 
$\bar{u} - \bar{d}$. 
This is found to be compatible with zero, but it is interesting to 
notice that the accuracy of this result is 
comparable to that of the valence distributions.

We conclude by recalling 
that the Sivers function can be disentangled from the transverse momentum convolution and 
extracted in a fully model-independent way (i.e., with no Gaussian assumption) by considering  
the asymmetries $A_h^w$ weighted with $P_{h \perp}$ \cite{Boer:1998}.  
The COMPASS Collaboration is currently working on the analysis of the weighted Sivers asymmetries.  
The method illustrated in the present paper can be applied 
to those observables in a straightforward way.

\begin{acknowledgments}
 
We are grateful to M.E.~Boglione 
for providing us the curves of the fits of Ref.~\cite{Anselmino:2012}.

\end{acknowledgments}

\end{document}